\documentclass[preprint,aps,superscriptaddress]{revtex4}
\usepackage{graphicx,epsfig,amssymb,amsmath, array}
\usepackage{relsize, float}
\usepackage[usenames]{color}

\begin{document}
\draft

%opening
\title{Bogdanov-Takens  resonance in time-delayed systems}

\author{Mattia Coccolo}
\affiliation{Nonlinear Dynamics, Chaos and Complex Systems Group, Departamento de F\'{i}sica,
Universidad Rey Juan Carlos, Tulip\'{a}n s/n, 28933 M\'{o}stoles (Madrid), Spain}
\author{BeiBei Zhu}
\affiliation{LSEC, ICMSEC, Academy of Mathematics and Systems Science, Chinese Academy of Sciences, Beijing 100190, China}
\affiliation{Departamento de Matem\'aticas, Universidad Carlos III de Madrid, Avenida de la Universidad 30, 28911 Legan\'es (Madrid), Spain}
\author{Miguel A.F. Sanju\'{a}n}
\affiliation{Nonlinear Dynamics, Chaos and Complex Systems Group, Departamento de F\'{i}sica,
Universidad Rey Juan Carlos, Tulip\'{a}n s/n, 28933 M\'{o}stoles (Madrid), Spain}
\affiliation{Institute for Physical Science and Technology, University of Maryland, College Park, Maryland 20742, USA}
\author{Jes\'{u}s M. Sanz-Serna}
\affiliation{Departamento de Matem\'aticas, Universidad Carlos III de Madrid, Avenida de la Universidad 30, 28911 Legan\'es (Madrid), Spain}

\date{\today}

\begin{abstract}

We analyze the oscillatory dynamics of a time-delayed dynamical system subjected to a periodic external forcing. We show that, for certain values of the delay, the response can be greatly enhanced by a {\it very small forcing amplitude}. This phenomenon is related to the presence of a \emph{Bogdanov-Takens} bifurcation and displays some analogies to other resonance phenomena, but also substantial differences.

 %\\
%{\bf keywords}:
%Nonlinear oscillations, Delay systems, Resonance,
\end{abstract}


\maketitle

\section{Introduction}

Different resonance phenomena play a key role in the sciences.  Examples, beyond the simplest case of a linear system forced at its natural frequency, include stochastic resonance \cite{gammaitoni,McDonnel}, chaotic resonance \cite{Zambrano}, coherence resonance \cite{Pikovsky} and vibrational resonance (VR) \cite{Landa}. For a recent monograph dealing with all these  phenomena, see \cite{Miguel}.
The stochastic resonance of a bistable system is triggered by the cooperation between noise and a weak periodic forcing, or even an aperiodic forcing. The noise can be replaced by a chaotic signal to obtain  chaotic resonance. It is also possible to have noise--induced resonance  in absence of  external periodic forces, a phenomenon called  coherence resonance. A nonlinear system driven by a biharmonic forcing,  with a frequency faster than the other, can show VR. Resonances appear not only in systems described by ordinary differential equations, but also in time-delayed systems. Time-delay effects arise frequently in practical problems and  have received  much attention in recent years \cite{Chiasson,Topics,Atay,Choe,Fischer}. Hereditary effects are sometimes unavoidable and may easily turn a well-behaved system into  one displaying very complex dynamics. A simple example is provided by Gumowski and Mira \cite{Gumowski}, who demonstrate that the presence of delays may destroy stability and cause periodic oscillations in  systems governed by differential equations. Vibrational resonance occurs in time-delayed systems with two harmonic forcings of different frequencies \cite{Yang1,Yang2,Jeevar,Fang}. Furthermore delay systems often possess oscillatory behavior even in the absence of forcing, and for this reason VR and related phenomena may occur even in the presence of only one external excitation \cite{Yang3,Yang4}.

In this work we present a new resonance phenomenon that may appear in systems with delay. The addition of a {\it very small external forcing} may result in the solution changing from a damped, small amplitude oscillation to a sustained, large amplitude oscillation. The sustained response takes place for a range of values of the frequency \(\Omega\) of the external forcing (as distinct from phenomena that require well-defined values of \(\Omega\)). The resonance occurs for a (narrow) interval of values of the delay and is related to the presence of a Bogdanov-Takens bifurcation \cite{kuznetsov} in the model. Therefore, we will refer to the resonance phenomenon as \emph{Bogdanov-Takens} resonance. The Bogdanov-Takens bifurcation of an equilibrium point appears in systems with two (or more) parameters when the equilibrium undergoing the bifurcation has a zero eigenvalue of algebraic multiplicity two \cite{Redmond,Gukenheimer}. Many different dynamics appears as explained in those references. In particular for some combinations of parameters values, one finds Hopf, homoclinic and periodic saddle node bifurcations near the Bogdanov-Takens bifurcation point.

\section{The system}

The model that we  use to describe and analyze the Bogdanov-Takens resonance is the apparently simple system called  delayed action oscillator \cite{Alvar}. It is a single variable system with a double-well potential and a linear delayed feedback term with a constant time delay $\tau\geq 0$. The oscillator can be written as:
\begin{align}
\label{eq:system}&\dot{x}=\alpha x_{\tau}+x-(1+\alpha)x^3+F\sin{\Omega t}\\
&x_{\tau}=x(t-\tau),
\end{align}
where $\alpha$ measures the influence of the returning signal relative to that of the local feedback and represents a negative feedback, $\tau$ is the time delay, and $F$ and $\Omega$ are the amplitude and frequency of the external periodic forcing. The constants  $\alpha$, $\tau$, $F$ and $\Omega$ are real and the interest is in the case \(\alpha\in(-1,0)\). Without the delayed term, this system would be a one-dimensional ODE and could not oscillate, but the linear delayed feedback converts the system into an infinite-dimensional one, allowing oscillatory dynamics. The system is interesting, among other things, for its analogy with the El Ni\~no Southern Oscillator (ENSO) \cite{Boutle,Krauskopf} and the well-known Duffing oscillator $\ddot{x}+\gamma\dot{x}+x(x^2-1)=0$, as discussed in \cite{Alvar}.

We begin by studying the unforced system with $F=0$ and parameters $\alpha, \tau$:
\begin{equation}\label{0Fsys}
\dot{x}=x+\alpha x_{\tau}-(1+\alpha)x^3.
\end{equation}
This has the equilibrium points $x=0$ and $x=\pm 1$. The equilibrium \(x=0\) is always unstable
as may be easily shown by studying the corresponding linearization of Eq.~(\ref{0Fsys}). The equilibria at \(\pm 1\) are stable in the absence of delay
 (\(\tau = 0\)), but undergo Hopf bifurcations \cite{junjie} in the delayed system. For $x=1$ (and for symmetry reasons for  $x=-1$),  the characteristic equation of the linearization is
\begin{equation}\label{character1}
\lambda=-3\alpha-2+\alpha e^{-\lambda \tau}.
\end{equation}
If $\alpha<-1$ or $\alpha>-1/2$, then, for any $\tau>0$, all roots of this equation  have negative real parts, and $x=1$ and  $x=-1$ are asymptotically stable.
If $-1<\alpha<-1/2$,  there is a sequence $\tau=\tau_k,~k=0,1,2,\dots$ of values of the delay for which
  Eq.~(\ref{character1})
 has a pair of imaginary roots $\pm i\omega_{0}$, where $\omega_0=\sqrt{\alpha^2-(3\alpha+2)^2}$. The delays \(\tau_k\) and the frequency \(\omega_0\) are related by the following expression:
\begin{equation}
\label{eq:relation_a_t}\tau_{k}=\frac{\sin^{-1}(-\omega_0/\alpha)+2k\pi}{\omega_0}.
\end{equation}
If $\tau\in [0,\tau_0)$, then all roots of Eq.~(\ref{character1}) have negative real parts.  For $\tau=\tau_0$, the roots of Eq.~(\ref{character1}) have  real parts \(<0\), except for the pair $\pm i\omega_0$. If $\tau\in (\tau_0,\tau_{1}]$, Eq.~(\ref{character1}) has  one pair of complex conjugate roots with positive real parts. Thus, for fixed $\alpha$, $-1<\alpha<-1/2$, and varying $\tau$, the equilibria $x=1$ and \(x=-1\) undergo a Hopf bifurcation at \(\tau=\tau_0\), where, as \(\tau\) increases, they turn from being asymptotically stable into being unstable. Additional Hopf bifurcations  occur at \(\tau_k\), \(k = 1, 2, \dots\), but we shall not be concerned with them. For the value \(\alpha = -0.925\)  used in~\cite{Alvar} and in the numerical experiments below, \(\tau_0\approx 1.1436\) and \(\omega_0\approx 0.5050\).

%Fig1
\begin{figure}[h]
   \centering
   \includegraphics[width=0.95\textwidth]{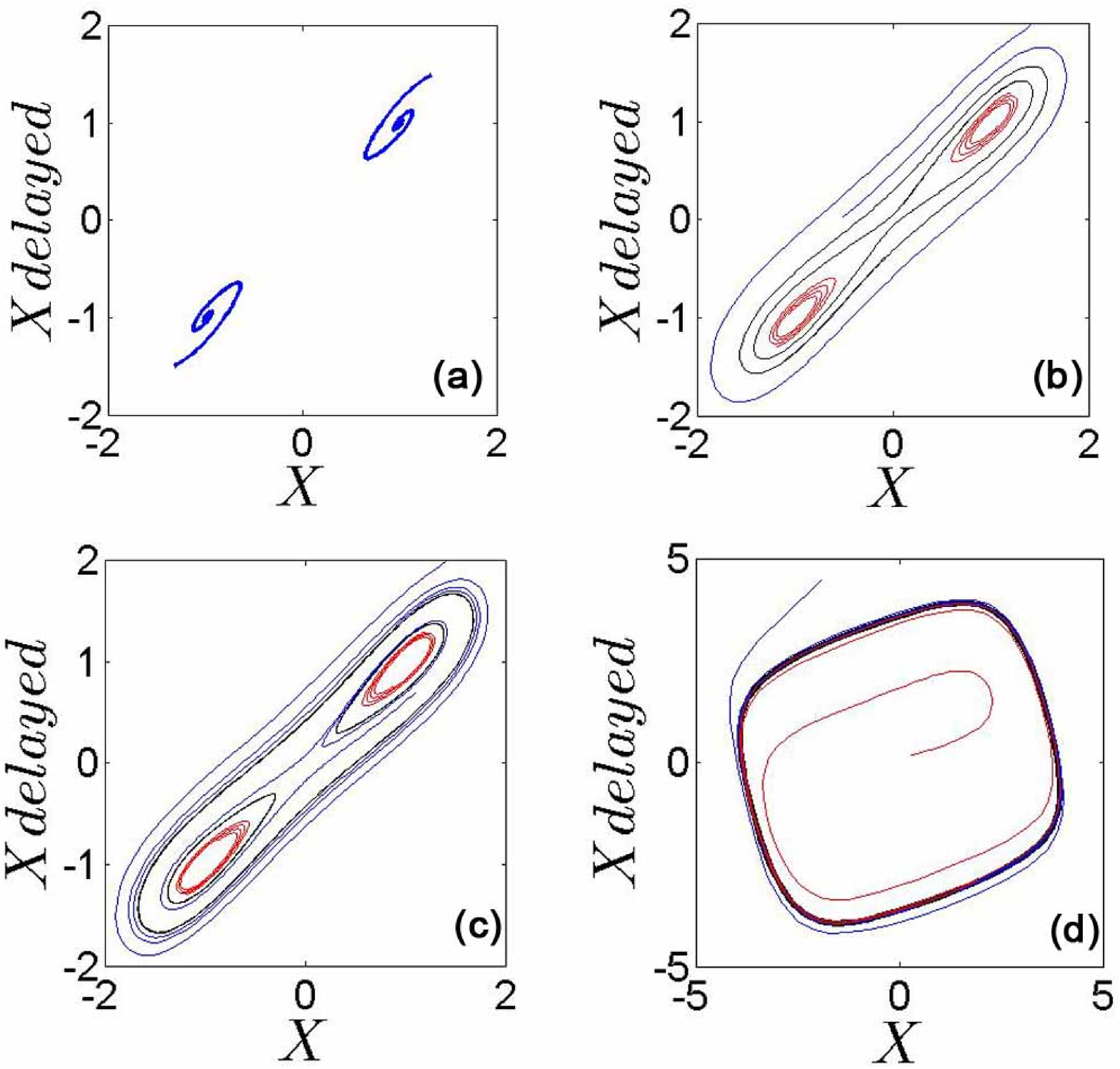}
   \caption{Phase portrait of the system, Eq.~(\ref{eq:system}), for $\alpha=-0.925$, and for $\tau=1, 1.122, 1.13$ and $1.7$, respectively. In the figures different trajectories  arising from different history functions, are plotted with different colours.  In panel (d) the black thick loop is the limit circle, to which the trajectories are attracted that we call stable loop, $L_s$. (colors available on line)}
\label{fig:phase_port_hist}
\end{figure}

\begin{table}[h]
\begin{center}
{\renewcommand{\arraystretch}{1}
\begin{tabular}{|m{3cm}|p{7cm}|}
\hline
\multicolumn{2}{|c|} {$\mathbf{ Solutions ~as ~a ~function ~of ~\tau ~for ~fixed ~\alpha. ~The ~Bogdanov-Takens ~bifurcation}$}\\
\hline
 $\tau<\tau_c$        & The equilibrium points \(\pm 1\) attract (most) solutions.\\
\hline
 \multicolumn{2}{|c|}{Periodic saddle--node bifurcation at $\tau_c$.}\\
 \multicolumn{2}{|c|} {A  stable loop \(L_s\) and a smaller unstable loop \(L_u\) are born.} \\
 \hline
 $\tau_c <\tau< \tau_h$   &  The equilibrium points attract solutions inside \(L_u\).\\
  & Outside \(L_u\) solutions are attracted to \(L_s\).\\
 \hline
 \multicolumn{2}{|c|} {Homoclinic bifurcation at $\tau=\tau_h$.}\\
  \multicolumn{2}{|c|}{The loop \(L_u\) gives rise to two unstable loops \(L_{\pm 1}\) around
 \(\pm 1\) respectively.}\\
 \hline
 $\tau_h<\tau<\tau_0$        & Solutions inside \(L_{\pm 1}\)  attracted to corresponding equilibrium. \\
 & Other solutions are attracted to \(L_s\).\\
 \hline
 \multicolumn{2}{|c|} {Hopf bifurcation at $\tau=\tau_0$.}\\
 \multicolumn{2}{|c|} {The loops \(L_{\pm 1}\) merge with the corresponding equilibrium.}\\
 \hline
Large $\tau$        &  \(L_s\) attracts most solutions.\\
 \hline
\end{tabular}}
\caption {Behaviour of the solutions of the unforced system \eqref{0Fsys} as a  function of $\tau$. $L_s$ represents the stable loop and $L_u$ the unstable loop, being both asymptotic trajectories of the system.} \label{tab:history_tau}
\end{center}
\end{table}

The panels in Fig.~\ref{fig:phase_port_hist} plot  \lq phase portraits\rq\ in the plane \(x,x_\tau\), showing the solutions of the system with fixed $\alpha$ and varying $\tau$. This behavior corresponds to a symmetric Bogdanov-Takens bifurcation as described in \cite{Redmond}. In each panel there are different solutions, plotted with different colors, corresponding to different constant history functions
\(x_\tau(t) = u_0\), \(t\in[-\tau,0]\), \(u_0\) a constant.
Bear in mind that this is different from a true phase portrait of an ODE system because in the delayed case it is not true that each point in the plane defines a unique trajectory. In the panels the solutions move counterclockwise. Panel (a) corresponds to the case of \lq\lq small\rq\rq\ \(\tau\); solutions are generically attracted to a stable equilibrum \(\pm 1\). For \(\tau\) \lq\lq large\rq\rq, solutions are generically attracted to a single big stable loop  that we call from now on  \(L_s\), see panel (d). As \(\tau\to \infty\), solutions on \(L_s\)  are  approximately  square waves where \(x(t)\) jumps from
$a= +\sqrt{ (1-\alpha) / (1+\alpha)}$ to $-a$ and back, and simultaneously \(x(t-\tau)\) jumps from $-a$ to $a$ and back.  The orbitally stable loop \(L_s\) is born at a saddle-node bifurcation at \(\tau=\tau_c\) (for $\alpha=-0.0925$,
$\tau_c\approx 1.119$). The saddle-node bifurcation point \(\tau_c\) is smaller than the Hopf bifurcation point \(\tau_0\) discussed above, so that for \(\tau\in(\tau_c,\tau_0)\) the attracting big loop \(L_s\) coexists with the attractors at \(x=\pm 1\). This is the regime of interest for our purposes. The interval \((\tau_c,\tau_0)\) contains  two subintervals \((\tau_c,\tau_h)\), \((\tau_h,\tau_0)\) corresponding to different dynamics. In the first of these subintervals (panel (b)), there is an unstable loop \(L_u\) surrounding the equilibria; \(L_u\) is of course born, together with \(L_s\), at the saddle-node bifurcation  at \(\tau=\tau_c\). At \(\tau=\tau_h\), \(L_u\)  becomes a homoclinic connection of the equilibrium at \(x=0\) and a further increase of \(\tau\) turns the homoclinic connection into  a couple of unstable orbits \(L_{\pm 1}\) , one around \(x=1\) and the other around \(-1\) (panel (c)). These unstable orbits disappear at the subcritical Hopf bifurcation at \(\tau=\tau_0\), where each of them merges with the corresponding equilibrium. The bifurcations at \(\tau_c\), \(\tau_h\) and \(\tau_0\) for fixed $\alpha$ clearly correspond to a Bogdanov-Takens scenario for the two-parameter model Eq.~(\ref{0Fsys}).
A summary of the possible behaviors of the solutions as $\tau$ varies appears in Table \ref{tab:history_tau}. Even though the results just reported here were obtained numerically, analytical calculations of the bifurcation diagram corresponding to Fig.~\ref{fig:phase_port_hist} can be found by means of the procedure presented in ~\cite{Redmond}.
There, the authors consider the general equation
\noindent
\begin{align}\label{eq:gen_eq}
&\dot{x}=f(x,x_{\tau})\\
&x_{\tau}=x(t-\tau),
\end{align}
where $\tau$ is the delay, and $f$ is an arbitrary smooth function.
They use the Taylor expansion in the right-hand side
\begin{equation}\label{eq:Ty_gen_eq}
\dot{x}=x+\alpha x_{\tau}+\gamma_1x^3+\gamma_2x^2x_{\tau}+\gamma_3xx_{\tau}^2+\gamma_4x_{\tau}^3+O(|x|^5)
\end{equation}
and study the linear stability and bifurcations conditions by calculating a centre manifold reduction. Then, they summarize the dynamics near the Bogdanov-Takens point $(\alpha,\tau)=(-1,1)$.
By following their steps, it is possible to find the centre manifold equations for our case $(\gamma_1=-(1+\alpha),\gamma_2=\gamma_3=\gamma_4=0)$, that yields
\begin{align}\label{eq:centreman}
&\dot{z_1}=z_2\\
&\dot{z_2}=(2\alpha+2)z_1+(\frac{-4\alpha}{3}+2\tau-\frac{10}{3})z_2+az_1^2z_2+2bz_1^3\\
&a=b=-2(1+\alpha).
\end{align}
For the $\alpha$ values that we have quoted in the previous section, the parameters $a,b$ are both negative. So that, our system fall in the second case discussed in detail in~\cite{Redmond}, where the conditions for the system to undergo the Bogdanov-Takens bifurcation in a small neighborhood of the Bogdanov-Takens point are explicitly given. These conditions reproduce our numerical results.

\section{Resonance}

\begin{figure}[http]
   \centering
   \includegraphics[width=0.85\textwidth]{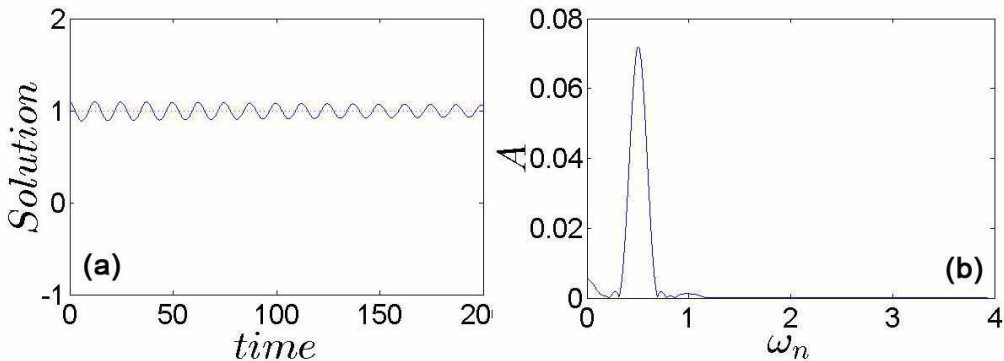}
   \caption{The unforced (\(F=0\)) system \eqref{eq:system} with $\alpha=-0.925$, $\tau=1.14$, and a constant history function $u_0 = 1.1$.    Panel
    (a): The solution \(x(t)\).  Panel (b): the Fourier analysis shows the frequency at which the amplitude peak shows up. The amplitude \(A\) has been calculated after omitting the initial transients.}
\label{fig:Sol_t114_F0_FFT}
\end{figure}

\begin{figure}[http]
   \centering
   \includegraphics[width=0.85\textwidth]{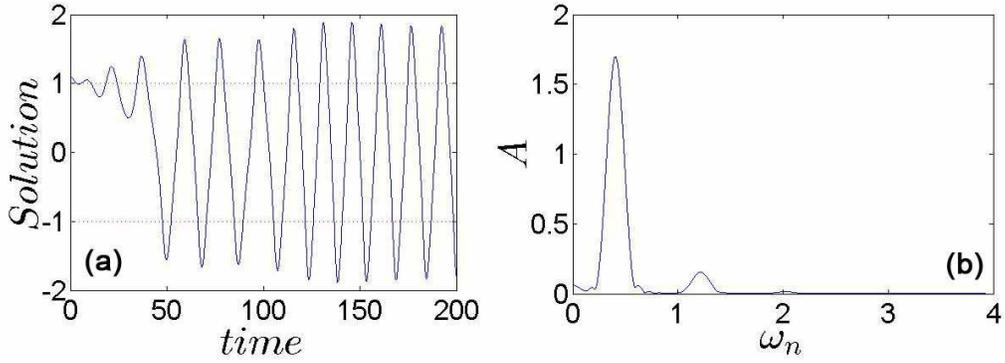}
   \caption{As in Fig.~\ref{fig:Sol_t114_F0_FFT}, except that a small forcing  $F=0.01$, $\Omega=0.5$ has been added; the solution is now a sustained oscillation of large amplitude. When comparing with Fig.~\ref{fig:Sol_t114_F0_FFT}(b) note the change in the vertical scale for \(A\).  The amplitude \(A\) has been calculated after omitting the initial transients.}
\label{fig:Sol_t114_F001_w05_FFT}
\end{figure}

\begin{figure}[http]
   \centering
   \includegraphics[width=0.85\textwidth]{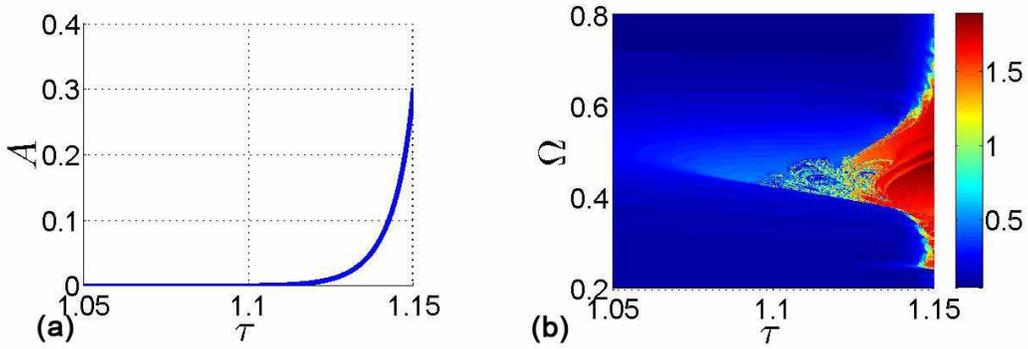}
   \caption{The figures show the amplitude gradient of the oscillatory solution of the system, Eq.~(\ref{eq:system}), as a function of $\tau$ for $F=0$ in panel (a), and as a function of $\Omega$ and $\tau$ for $F=0.01$ in panel (b). We have considered values of $\tau$  close to the critical values where the bifurcations occur. It is possible to appreciate the complexity of the panel (b) for values around $\Omega=0.5$. Here, $\alpha=-0.925$ and the history function is the constant $u_0 = 1.1$. The amplitude \(A\) has been calculated after omitting the initial transients. (colors available on line)}
\label{fig:Ampl_t_F0_F001_w05_a0925}
\end{figure}

The forced system $\dot{x} = x-x^3+ A \cos\omega t$ shows small-amplitude sustained oscillations around one of the equilibrium points $x=\pm1$, which are only possible due to the slow forcing $A\cos\omega t$. Then, by adding a fast forcing $B\cos{\Omega t}$, with $\Omega \gg \omega$, the oscillations may go from one well to the other. This is the phenomenon of vibrational resonance \cite{Landa,ander}. Equations like (\ref{eq:system})  may exhibit something extremely similar~\cite{Alvar_bio,Yang4}. The autonomous system $\dot{x}+x ((1+\alpha)x^2-1) - \alpha x_\tau = 0$ shows slowly damped oscillations around $+1$ or $-1$, induced by the delay. Then, the addition of a forcing term $F \sin\Omega t$ may give rise to sustained oscillations that go from one well to the other. The phenomenon that we study here is considerably different. We illustrate it in the case  with $\alpha = -0.925$, $\tau = 1.14$ and constant history $u_0= 1.1$. For this value of \(\tau\), the equilibria \(\pm 1\) coexist with the attractor \(L_s\). Figure~\ref{fig:Sol_t114_F0_FFT} corresponds to the unforced case \(F=0\).
The solution is a marginally damped oscillation with angular frequency approximately equal to $\omega_n=0.50$, as the position of the peak in Fig.~\ref{fig:Sol_t114_F0_FFT}(b) shows.
Then, we add \emph{a very small forcing value} $F = 0.01$ of angular frequency $\Omega = 0.50$ (the exact value of the forcing frequency $\Omega$ is not critical, as we will discuss later). As we show in Fig.~\ref{fig:Sol_t114_F001_w05_FFT}(a),  the solution is a sustained oscillation of large amplitude and  angular frequency $\omega_n=0.40$, as shown by the position of the peak in Fig.~\ref{fig:Sol_t114_F001_w05_FFT}(b). Therefore, there is a huge impact of the small forcing term. The resulting sustained oscillation is triggered by the forcing, but it is not a direct response to it, because the frequency of the interwell oscillation does not match the forcing frequency $\Omega$, as we  see by comparing Figs.~\ref{fig:Sol_t114_F0_FFT}(b) and \ref{fig:Sol_t114_F001_w05_FFT}(b). In a phase portrait, the forcing would cause the solution to jump from the neighbourhood of the equilibrium \(x=1\) to the stable loop \(L_s\).
Figure~\ref{fig:Ampl_t_F0_F001_w05_a0925} shows the amplitude of the solution as a function of $\tau$,  without forcing  (panel (a)) and as a function of $\tau$ and $\Omega$, with $F=0.01$ (panel (b)). The numerical experiments support the analysis done in the previous section. In fact, it is possible to appreciate the enhancement of the amplitude $A$ for \(\tau\) in panel (a), and the enhancement of the amplitude $A$ in the range $1.119<\tau<1.143$ where \(L_s\) coexists with the stable equilibria,  in panel (b). If $\tau$ is larger than $1.143$, then the equilibrium points $\pm 1$ loose their stability so that the damped oscillations around them no longer exist, and the system shows an interwell oscillation without any need of an external forcing. On the other hand, if $\tau$ is below $1.119$, the solution will eventually settle in one of the wells, even if in a transient phase it oscillates between both wells. Moreover, by looking at the $\Omega$-axis it is possible to appreciate the complexity of the amplitude values around the value of $\Omega=0.5$ due to the nonlinearity of the system, as predicted in the previous paragraphs. It is worth to point out that the resonance phenomenon takes places for values of $\Omega$ in a suitable interval, rather than at critical values.
%Fig2
%Fig3

%Fig4

It is important to point out that the phenomenon that we are discussing is very different from well-known cases where a forcing with a moderate value of $F$ gives a solution that, upon Fourier analysis, is seen to consist of modes $\cos(\Omega t+ \phi_1)$ (the fundamental harmonic),  $\cos(3 \Omega t+ \phi_3)$ (the third harmonic) or $\cos(\Omega t/3 + \phi_1)$ (subharmonic) (odd numbered overtones are expected in view of the cubic nonlinearity).  Similarly, the phenomenon here is clearly different form that described in \cite{Yang3}.

\begin{figure}[http]
   \centering
   \includegraphics[width=0.85\textwidth]{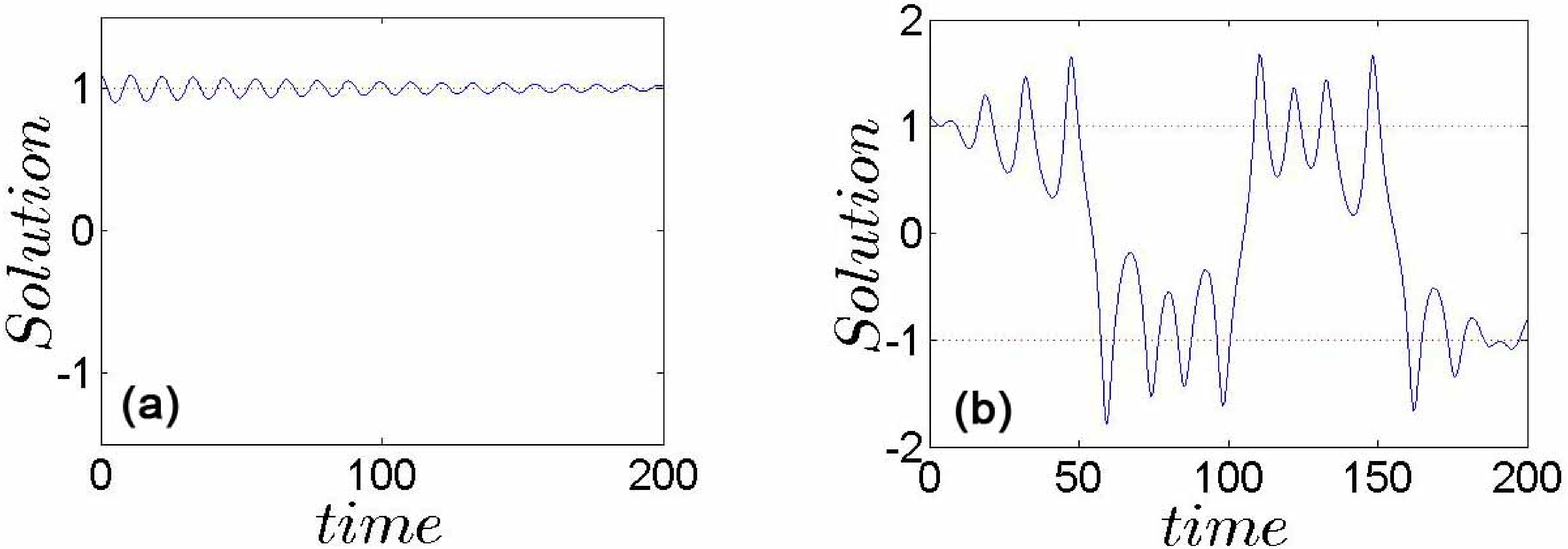}%new figure by beibei
   \caption{Model \eqref{eq:system_ext} with  $\tau=1.19$, $\alpha=-0.9$, and a constant history function $u_0=1.1$.
   On the left: damped oscillations in the absence of forcing (\(F=0\)). On the right: large amplitude oscillations
   for a small forcing, $F=0.015$, $\Omega=0.45$. }
\label{fig:ampl_sys_ext}
\end{figure}

 In order to show that the phenomenon is not specific to the particular model \eqref{eq:system}, we have also analyzed the equation
 \begin{align}
\label{eq:system_ext}&\dot{x}=\alpha x_{\tau}+x-3(1+\alpha)xx_{\tau}^2+2(1+\alpha)x_{\tau}^3+F\sin{\Omega t}\\
&x_{\tau}=x(t-\tau),
\end{align}
that undergoes a Bogdanov-Takens bifurcation \cite{Redmond}.
The resonance studied here also occurs, as seen in Fig.~\ref{fig:ampl_sys_ext}: the introduction of a {\it very small external forcing} induces again interwell oscillations.
%fig5

\section{Dynamics of the resonance}
We now study the impact on the resonance of changes in the forcing frequency  $\Omega$, the parameter $\alpha$ and the history function $u_0$.
%Fig6
\begin{figure}[http]
   \centering
   \includegraphics[width=0.85\textwidth]{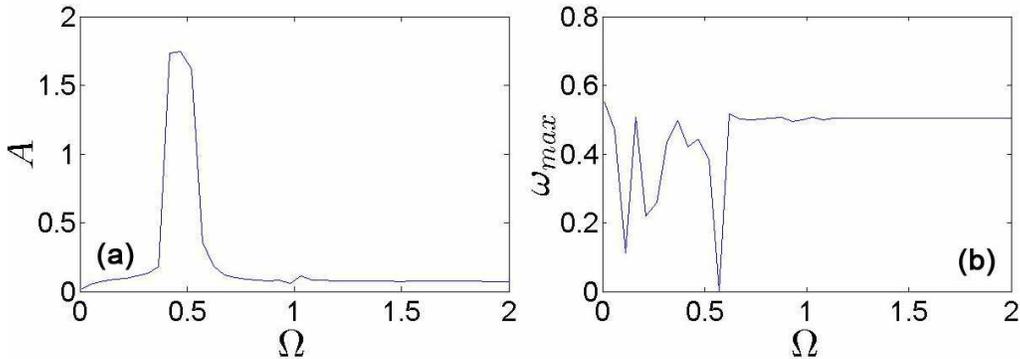}
   \caption{The figures show (a) the amplitude of the oscillation and (b) the frequency $\omega_n$ which gives the maximum in amplitude, that we called $\omega_{max}$,  as a function of $\Omega$, corresponding to the system, Eq.~(\ref{eq:system}), for $F=0.01$, $\tau=1.14$, $\alpha=-0.925$, and a constant history function $u_0=1.1$. In (a) a well-defined peak appears  around $\Omega=0.4$ that shows  that the resonance phenomenon takes place for suitable values of the forcing frequency. The amplitude \(A\) has been calculated after omitting the initial transients.}
\label{fig:ampl_w_F001_t114_a0925}
\end{figure}

Figure~\ref{fig:ampl_w_F001_t114_a0925}(a) depicts the amplitude  (maximum value of the Fourier spectrum) of the response
\(x\) as a function of \(\Omega\); the resonance manifests itself for a range of values of \(\Omega\) around 0.4. Panel (b) gives the frequency \(\omega_n\) for which the Fourier spectrum of the signal attains its maximum value,  that we called in the figure $\omega_{max}$. Note the little correlation between \(\omega_{max}\) and \(\Omega\); for \(\Omega\) large,  \(\omega_{max}\) corresponds to the frequency on \(L_s\).

\begin{figure}[http]
   \centering
   \includegraphics[width=0.85\textwidth]{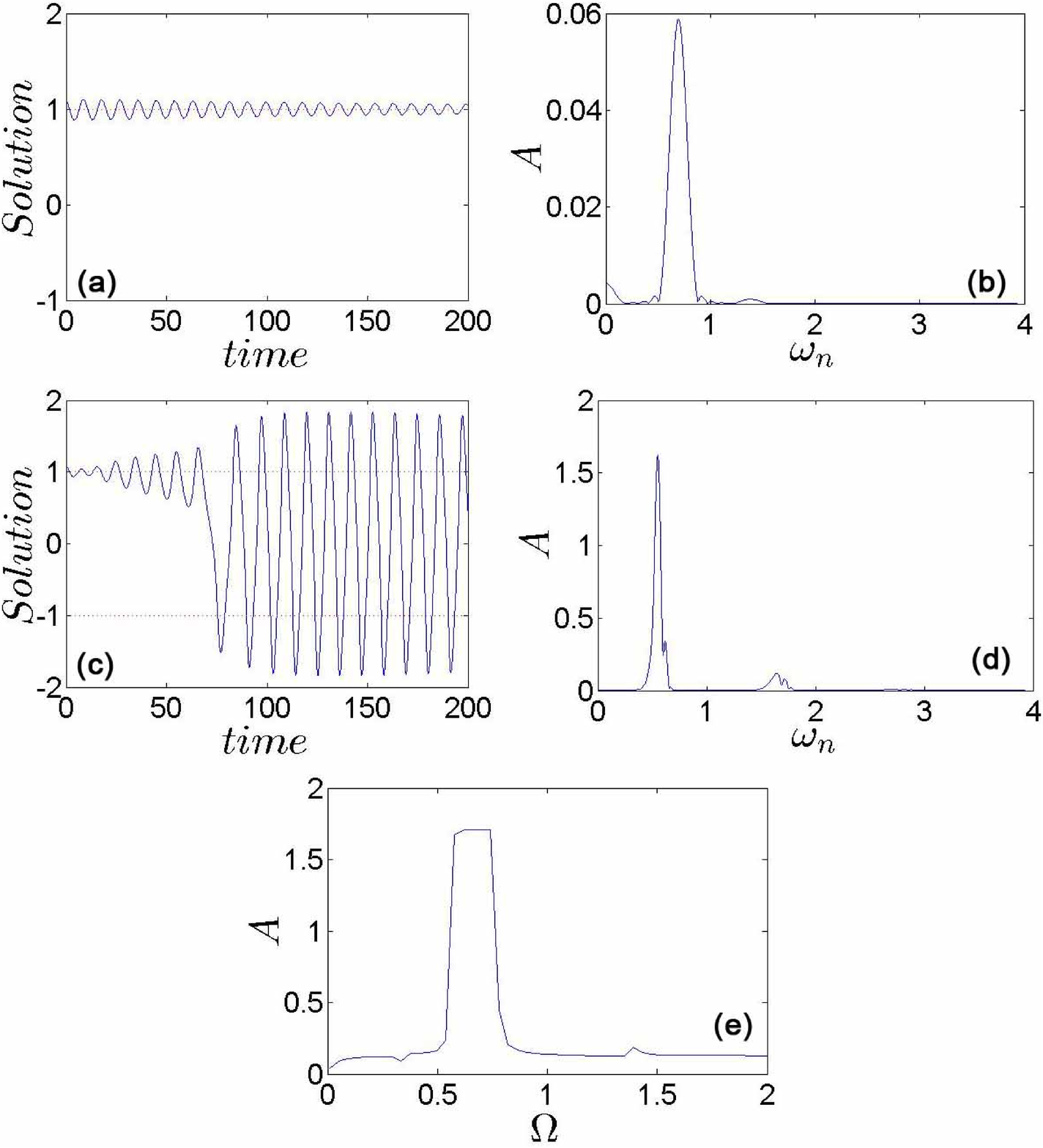}
   \caption{The resonance for Eq.~(\ref{eq:system}), but now with $\alpha=-0.8$, $\tau=1.5$. Panels (a) and (b) show  the solution and its Fourier analysis for $F=0$. Panels (c) and (d) show  the solution and its Fourier analysis for $F=0.01$, $\Omega=0.6$. Panel (e) shows the amplitude of the solution as a function of $\Omega$. The amplitude \(A\) has been calculated after omitting the initial transients. }
\label{fig:am08_panel}
\end{figure}

In Fig.~\ref{fig:am08_panel} we use the alternative value $\alpha=-0.8$  in order to check  the occurrence of the resonance. Note that the value of the critical $\tau_0$ for which the resonance appears increases, in agreement with Eq.~(\ref{eq:relation_a_t}).

%Fig7

%fig8
\begin{figure}[http]
   \centering
   \includegraphics[width=0.85\textwidth]{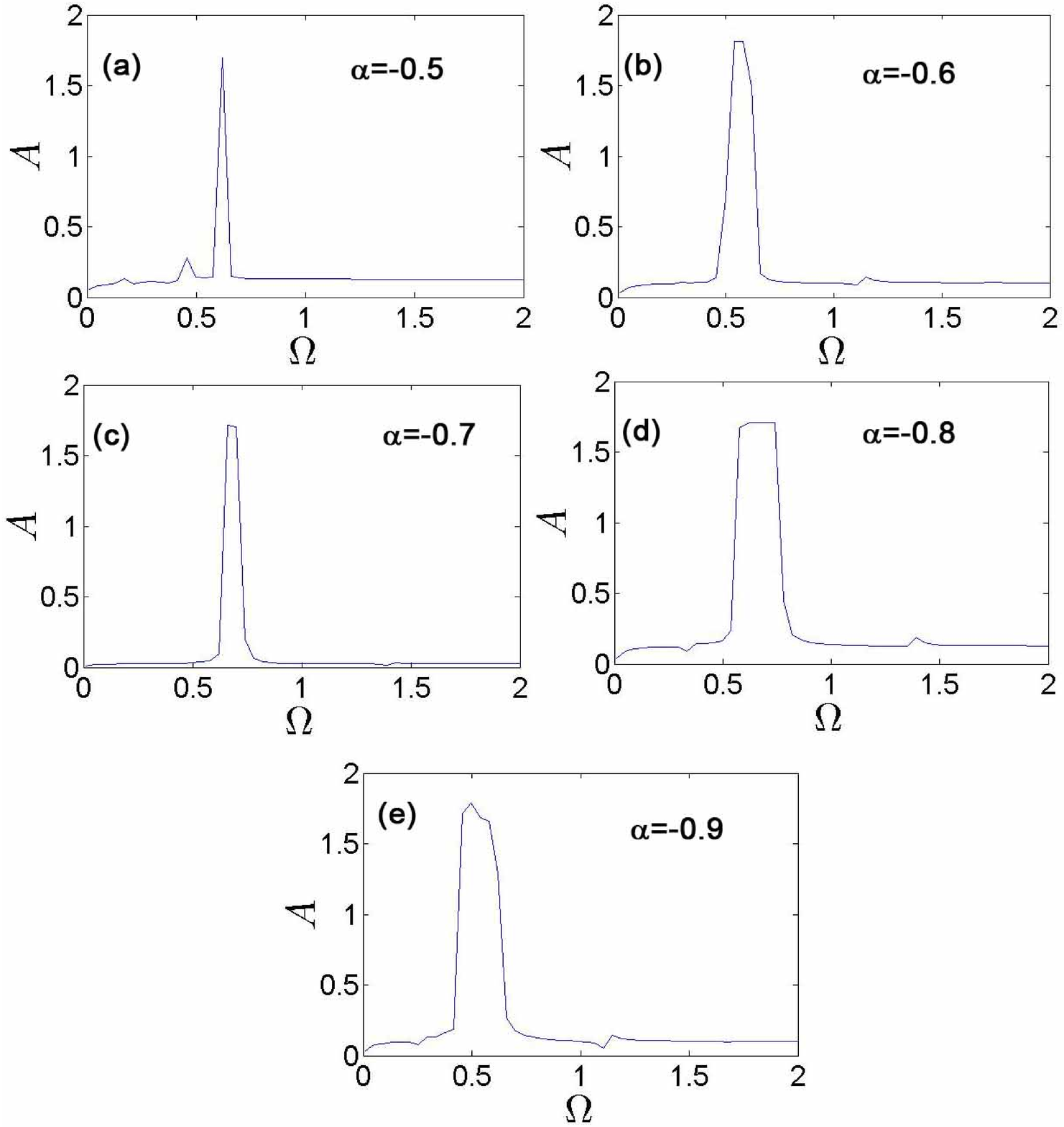}
   \caption{Amplitude of the solution of  Eq.~(\ref{eq:system}) as a function of $\Omega$ for different values of the parameter $\alpha$. Panel (a):  $\tau=12.8$ and $\alpha=-0.5$. Panel (b):  $\tau=3.3$ and $\alpha=-0.6$. Panel (c):  $\tau=2.01$ and $\alpha=-0.7$. Panel (d): $\tau=1.51$ and $\alpha=-0.8$. Panel (e):  $\tau=1.2$ and $\alpha=-0.9$. As $\alpha$ decreases, the values of $\tau$ that trigger the  resonance decrease. The amplitude \(A\) has been calculated after omitting the initial transients.}
\label{fig:ampl_w_alphas}
\end{figure}

It is of interest to study numerically the changes of the critical $\tau_0$ as $\alpha$ varies, as shown in Fig.~\ref{fig:ampl_w_alphas}, that plots
the amplitude of the oscillation as a function of $\Omega$ for different values of $\alpha$, from $\alpha=-0.5$, panel (a) to $\alpha=-0.9$, panel (e). We note that for larger values of $\alpha$, the critical value of $\tau$ that triggers the resonance increases. The figures also show that the shape of the peak in  the \( (\Omega,A) \) plane and the range of $\Omega$ leading to resonance change with \(\alpha\), although not as much as the value of $\tau_0$.

Another important factor in the study of delayed systems is the history function. We have carried out  numerical experiments  changing the history function and  found that the phenomenon is robust against the variation of the history. In fact, none of the figures shown above  changes  if we alternatively use linear, quadratic or sinusoidal histories.

\section{Conclusions}

In conclusion, we have shown the phenomenon of  {\it Bogdanov-Takens resonance} in time-delayed systems. This resonance is produced when a periodic signal of a very small amplitude is applied to a delayed system that undergoes, for some parameter values, a Bogdanov-Takens bifurcation. This means that the forcing causes the solution to jump from the unstable loop $L_u$ in the neighbourhood of the equilibrium, to the stable loop $L_s$, due to the coexistence of these two unstable and stable solutions in the formerly mentioned bifurcation. Furthermore, the frequency of the resulting sustained oscillation is not related to the frequency \(\Omega\) of the forcing. Resonance takes places for \(\Omega\) in a suitable interval, rather than at critical values of \(\Omega\).
%periodic signal of a very small amplitude applied to a delayed system produces sustained oscillations of large amplitude. The frequency of the resulting sustained oscillation is not related to the frequency \(\Omega\) of the forcing. Resonance takes places for \(\Omega\) in a suitable interval, rather than at critical values of \(\Omega\).

\section*{Acknowledgements}

The work of M. C. and M. A. F. S.  was supported by the Spanish Ministry of Economy and Competitiveness under project no. FIS2013-40653-P and by the Spanish State Research Agency (AEI) and the European Regional Development Fund (FEDER) under project no. FIS2016-76883-P. In addition, M. A. F. S. acknowledges the jointly sponsored financial support by the Fulbright Programme and the Spanish Ministry of Education (programme no. FMECD-ST-2016). B. Z. has been supported by the National Natural Science Foundation of China (Grant No. 11371357). She is grateful to Universidad Carlos III de Madrid for hosting her stay in Spain and to the China Scholarship Council for providing the necessary funds. J. M. S. has been supported by projects MTM2013--46553--C3--1--P and MTM2016--77660--P(AEI/FEDER, UE) from  the Spanish Ministry of Economy and Competitiveness.

\end{document}